# Study of QCD generalized ghost dark energy in FRW universe

Mahasweta Biswas[1,a], Ujjal Debnath[1,b], Shounak Ghosh[2,c] , B. K. Guha[2,d]

[1] Department of Mathematics, Indian Institute of Engineering Science and Technology, Shibpur, Howrah, West Bengal 711103, India
[2] Department of Physics, Indian Institute of Engineering Science and Technology, Shibpur, Howrah, West Bengal 711103, India

**Abstract** A phenomenological generalized ghost dark energy model has been studied under the framework of FRW universe. In ghost dark energy model the energy density depends linearly on Hubble parameter (H) but in this dark energy model, the energy density contains a the sub-leading term which is depends on $\mathcal{O}(H^2)$, so the energy density takes the form $\rho_D = \alpha H + \beta H^2$, where $\alpha$ and $\beta$ are the constants. The solutions of the Friedman equation of our model leads to a stable universe. We have fitted our model with the present observational data including Stern data set. With the help of best fit results we find the adiabatic sound speed remains positive throughout the cosmic evolution, that claims the stability of the model. The flipping of the signature of deceleration parameter at the value of scale factor $a = 0.5$ indicates that the universe is at the stage of acceleration i.e. de Sitter phase of the universe at late time. Our model shows that the acceleration of the universe begin at redshift $z_{ace} \approx 0.617$ and the model is also consistent with the current observational data.

## 1 Introduction

The success of FLRW model of the universe and observation results from Hubble's law has proved that our universe is expanding with an acceleration. First direct observational evidence in favour of this was identified by analysing the outburst of Type Ia supernova in 1998 [1–4]. Nowadays lot of evidences are available to prove the accelerating phase of present Universe.

To explain this current phase of the Universe under the framework of Einstein's gravity, an energy has to be considered which is responsible for this acceleration. In literature this energy has been termed as dark energy (DE). This type of energy having negative pressure creates a repulsive force acting radially outward that causes the expansion of the universe. Though it has not been detected yet but various observational evidences confirm the existence of DE. The equation of state in connection with the isotropic pressure and matter density can be represented as $p = \omega\rho$, where $\omega$ is the EOS parameter. For DE $\omega$ has the value as, $\omega < -\frac{1}{3}$ i.e., $\frac{p}{\rho} < -\frac{1}{3}$ [1–4]. So one can start investigation on expansion of the universe with an acceleration adopting any new DE model if and only if the EoS parameter satisfies the condition $\omega_D < -\frac{1}{3}$ (here D stands for DE).

It is an important challenge to the researcher to find the nature of this DE. One of the most known dark energy candidate is the Einstein's Cosmological Constant ($\Lambda$). The EOS parameter ($\omega_D$) for this dark energy candidate has the value $-1$. At the time of construction of the general theory of relativity (GR) by Einstein, it has been found that the solutions of Einstein's field equations showed an expanding nature of the universe. In order to stabilize the universe Einstein had introduced this cosmological constant in his field equations that made the solutions static. Later the observational evidence and success of Hubble's law made Einstein to realise that the universe is in the phase of expansion. So he discarded this cosmological constant from his field equation. But the cosmological constant than made several come backs in various situations to explain this phase of the universe. In order to explain one loop quantum fluctuations i.e., Casimir effect $\Lambda$ plays an important role. In the year 1998, after the results of supernova explosions it was established that the universe is expanding with an acceleration [1,2,5,6] and the cosmological constant got a permanent place in literature. At present this $\Lambda$ has been considered as one of the most responsible candidate for this phase of acceleration of the universe. Inspite of it's success as a candidate of dark energy, it has been suffering from two major drawbacks such as fine tuning problem and coincidence problem. Apart from Einstein's cosmological constant varieties of dark energy scenarios have been discussed in literature for the explanation of the accelerating phase of the present universe. One can find the review on dark energies in Refs. [7–10].

[a] e-mail: mahasweta.rs2014@math.iiests.ac.in
[b] e-mail: ujjaldebnath@gmail.com
[c] e-mail: shounak.rs2015@physics.iiests.ac.in
[d] e-mail: bkguhaphys@gmail.com

To explain this accelerating phase of present universe by these recent models of dark energy, generally we require a new degree of freedom [11–16]. Any new dark energy model has many hidden features and can give rise to a new challenge to the researchers. However it would be very much desirable, if one can resolve the DE problem avoiding the choice of extra degree of freedom or any extra parameter, as these may create inconsistencies. One of the most appealing model under this category is the so called " ghost dark energy" (GDE) or Veneziano ghost dark energy [17–21]. The term "ghost" is interpreted in the cosmology community in the conventional context as a propagating degree of freedom with the negative norm (such as the "rolling ghost" model, "ghost condensation" model, etc). Veneziano ghost is the main ingredient of this model and Veneziano ghost field is responsible for the recent cosmic acceleration. This Veneziano ghost field was introduced in literature to resolve the $U(1)$ problem in low energy compelling theory of Quantum Choromodynamics (QCD) [22–27]. Though it is very abstract in the Minkwoski spacetime quantum field theory, it shows some essential physical effects in dynamic spacetime or spacetime with nontrivial topology. Nothing is contributed to the vacuum energy density in the flat Minkwoski spacetime by this ghost dark energy, but in non flat system the ghost dark energy creates a small vacuum energy density. This vacuum energy density in the non flat spacetime is directly proportional to $\Lambda_{QCD}^3 H$, where $\Lambda_{QCD}^3$ is represent the QCD mass scale [28] and $H$ is the Hubble parameter. The dynamical cosmological constant [17–21] has been considered to explain the vacuum energy density of the ghost field in GDE model. Various features of GDE can be found in details in literatures [29–32]. Taking $\Lambda_{QCD} \sim 100\ Mev$ and $H \sim 10^{-33}\ eV$ one can obtain the magnitude of $\Lambda_{QCD}^3 H$, as approximately $(3 \times 10^{-3} ev)^4$ for the density of observed dark energy [21]. Based on these results lot of works have been done on GDE.

Sheyki and Sadegh [33] have explored the characteristics of GDE in non-flat universe in presence of the interaction between the dark energy and dark matter. The phenomenological significance of GDE has been studied by Cai et al. [32]. Many other features of GDE model have been observed in the Refs. [29–31,33–38].

From the above references the energy density of ghost dark energy can be represented as

$$\rho_D = \alpha H, \tag{1}$$

Here $H$ is Hubble parameter and $\alpha$ is a constant with dimension of $[energy]^3$. $H$ can be expressed as $H = \frac{\dot{a}}{a}$, where $a$ is the scale factor and dot represent the time derivative of $a$.

In Ref. [39], the author claims that contribution to vacuum energy density by the Veneziano ghost dark energy does not exactly depends on $H$ but a sub-leading term of $H^2$ must appear in the expression of energy density. This sub-leading term plays an important role for evolution of the universe at the early stage. It might also represent the early DE [40,41]. With this sub-leading item, this GDE model is called the generalized ghost dark energy (GGDE) model and the energy density can be represented as

$$\rho_D = \alpha H + \beta H^2, \tag{2}$$

where $\beta$ is another constant having the dimension of $[energy]^2$. The Veneziano ghost in GGDE model does not violate unitarity, causality, gauge invariance and other important features of renormalizable quantum field theory [42]. From the observational evidences and data analysis, Cai et al. [40] have showed that the GGDE model is more acceptable and more accurate than the usual GDE model. This results attracted the interest of researcher to work on GGDE model in recent days [41,43,44].

In GDE model Cai et al. [40] claims the stability of the model without satisfying the condition of sound speed. In addition to it, many other references are also available in literature where it has been found that the squared speed of sound remains negative for GDE model. The study of the squared sound speed ($v_s^2 = \frac{dp}{d\rho}$) is one of the most important parameter to investigate the stability of any model. The sign of $v_s^2$ plays crucial role for determining stability. If $v_s^2 < 0$ the solution become classically unstable against perturbation. This issue has been studied for many DE candidates. It was observed that $v_s^2$ is positive for Tachyon and Chaplygin gas DE models which demands the stability of these models against perturbation [45,46]. However the perfect fluid of holographic DE with future event horizon and agegraphic model of DE is found unstable against perturbation as $v_s^2$ is negative [47,48]. Again Ebrahimi et al. [31] have showed that the QCD ghost dark energy model is unstable against perturbation due to negative value of squared sound speed (i.e., $v_s^2 < 0$). The instability caused due to the treatment of ghost as a propagating physical degree of freedom. So they concluded that the components of QCD ghost dark energy and DM cannot provide a stable solution of the universe. Whereas the above result will not hold for GGDE model as it has not any propagating degrees of freedom [42]. Similar feature is also holds for pure de Sitter space (equation of state $p = -\rho$). Now following the naive computations one can found the sound speed as $v_s^2 = dp/d\rho = -1$. This result is not a signal for instability as there is no any propagating degree of freedom in pure de Sitter case. That indicates that pure de Sitter vacuum is obviously stable solution. The computation of the speed of sound makes no sense in this case. Whereas considering the GGDE model with proper choice of the value of $\beta$ we have obtained a stable solution of the universe by analysing various physical properties. We get the value of sound speed within 1 ($0 < v_s^2 < 1$) throughout the stellar evolution. Following the argument of Ebrahimi and Sheykhi [31], it can also be claimed that our model must be classically stable under perturbations.

So motivating with the works on GGDE we have studied the model under the FRW background of a flat homogenous and isotropic universe. In this work we have discussed the dynamics of GGDE and evaluate the solution of Friedman equation in Sect. 2. In this section we also studied various physical properties of GGDE such as cosmic time evolution, energy density, EoS parameter. We have shown the results of data fitting. Using these results we have analysed the classical stability in Sect. 3, which is followed by physical results and conclusion Sects. 4 and 5 respectively.

## 2 Dynamics of generalised ghost dark energy

To study the dynamics of generalised ghost dark energy (GGDE) we have considered the Friedman–Robertson–Walker (FRW) universe. We are restricting ourselves only in two components of the energy viz., CDM and DE, neglecting the contributions for the radiation and baryons which are very small in comparison with the other terms. So we have the Friedman equation as,

$$H^2 = \frac{8\pi G}{3}(\rho_D + \rho_{DM}), \tag{3}$$

where $\rho_D$ represents energy density for generalised ghost dark energy and $\rho_{DM}$ that for the dark matter. Inserting the expression of $\rho_D$ from Eq. (2) in Eq. (3), we have left with

$$H^2 = \frac{8\pi G}{3}(\alpha H + \beta H^2 + \rho_{DM}). \tag{4}$$

Now the continuity equation for $\rho_{DM}$ can be written in the following form

$$\dot{\rho}_{DM} + 3H\rho_{DM} = 0 \Rightarrow \rho_{DM} = \rho_{DM0}a^{-3}. \tag{5}$$

For the present time we take $a_0 = 1$, where the subscript zero stands for representing the value of scale factor at present time.

From Eqs. (4) and (5), one can obtain the Raychaudhuri equation as

$$\dot{H} + H^2 = -\frac{4\pi G}{3}\left[-\rho_D\left(\frac{\dot{\rho}_D}{H\rho_D} + 2\right) + \rho_{DM}\right]. \tag{6}$$

Now solving $H$ from Friedman equation (4) eventually we have

$$H_{\pm} = \frac{\frac{8\pi G\alpha}{3} \pm \sqrt{\frac{8\pi G\alpha}{3}^2 + \frac{32\pi G}{3}\gamma\rho_{DM0}a^{-3}}}{2\gamma}. \tag{7}$$

where $\gamma = 1 - \frac{8\pi G\beta}{3}$. Here we have two values of $H$. $H_+$ denotes expansion of the universe whereas $H_-$ represents the contraction. Based on the observational results we can ignore the later one. So we have left with $H_+$ only. For simplicity we use $H$ instead of $H_+$ throughout our study.

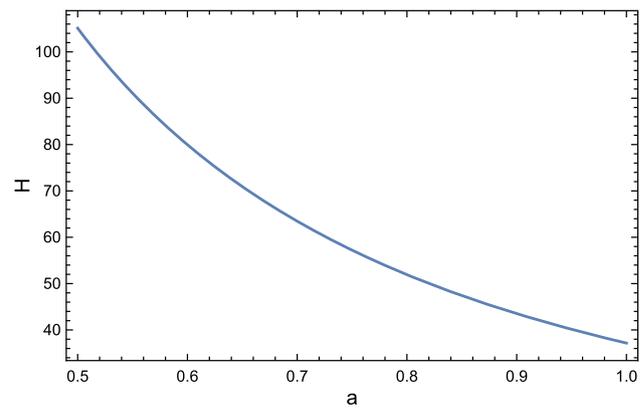

**Fig. 1** Variation of $H$ with $a$

$$H_+ = H = \frac{\frac{8\pi G\alpha}{3} + \sqrt{\frac{8\pi G\alpha}{3}^2 + \frac{32\pi G}{3}\gamma\rho_{DM0}a^{-3}}}{2\gamma}. \tag{8}$$

Variation of Hubble parameter with respect to scale factor '$a$' is shown in Fig. 1. From the figure it is clear that value of Hubble parameter is decreases with the evolution of the Universe.

Before going to the next step one can notice that Eq. (4) exactly matches with the Dvali-Gabadadze-Porrati (DGP) model with $r_c \sim \frac{1}{\alpha}$, though generalised ghost DE model and DGP model are completely different. It can be observed that for a flat universe the Friedman equation is same for both the models, whereas this equation is different for a non-flat universe having spacial curvature $\kappa$ ($\kappa \neq 1$). Again, GGDE model is more faithful theory than DGP model because DGP model is five dimensional one in the brane world scenario rather GGDE is a four dimensional model and it is the domain of general relativity.

We have the Friedman equation for generalised ghost DE model in a non-flat FRW universe as

$$H^2 + \frac{k}{a^2} = \frac{8\pi G}{3}(\alpha H + \beta H^2 + \rho_{DM}), \tag{9}$$

while that of in DGP model can be written as

$$H^2 + \frac{k}{a^2} - \frac{1}{r_{cs}}\sqrt{H^2 + \frac{k}{a^2}} = \frac{8\pi G}{3}(\rho_D + \rho_{DM}). \tag{10}$$

Now to continue our present discussion with generalised ghost dark energy, let us define a characterised scale factor denoted by $a_\star$ as

$$a_\star \equiv \left(\frac{12\rho_{DM0}\gamma}{8\pi G\alpha^2}\right)^{\frac{1}{3}}$$

$$= \left(\frac{4\Omega_{m0}\gamma(\alpha + \beta H_0)^2}{\Omega_{D0}\alpha^2}\right)^{\frac{1}{3}}, \tag{11}$$

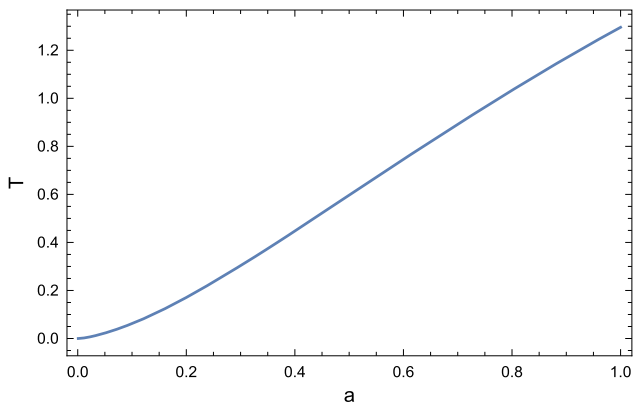

**Fig. 2** Variation of cosmic time evolution $T \; (= \frac{4\pi G\alpha(t-t_i)}{\gamma})$ with $a$

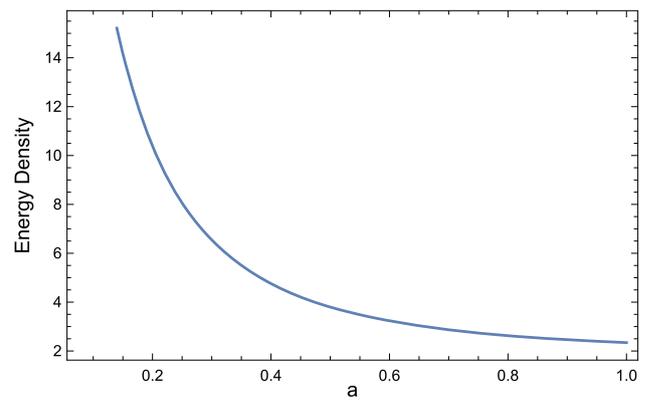

**Fig. 3** Variation of energy density ($\frac{\rho_D(3-8\pi G\beta)}{4\pi G\alpha^2}$) with $a$

where $\Omega_{m0}$ and $\Omega_{D0}$ are the dimensionless parameter and respectively represents the energy density of CDM and dark energy. This characteristics scale factor actually represents the transition point of the universe from the dust phase to present de Sitter phase. If we assume the values of $\Omega_{m0}$ and $\Omega_{D0}$ to be 0.2034 and 0.7966 respectively then we approximately get $a_\star \sim 1$, this indicate that the transition takes place just at present. In order to study the variations of various parameters we have expressed $\alpha$ and $\beta$ in Planck scale, both the parameters become dimensionless in Planck scale. So in our article we have taken only the values of $\alpha$ and $\beta$.

### 2.1 Cosmic time evolution

Now from Eq. (8) we can observe that for $a \ll a_\star$, i.e at early time of the universe, $H \propto a^{-\frac{3}{2}}$, this indicates the dust phase of the universe. For late time at $a \gg a_\star$ Eq. (8) indicates that $H =$ constant, denotes the entry at the de Sitter phase at the later epoch. As mentioned earlier $a_\star$ represents the transition point between the two epochs. We can solve the Eq. (8) analytically and we get

$$\frac{4\pi G\alpha(t-t_i)}{\gamma} = -y^3 + y^3\sqrt{1+y^{-3}} + \frac{3}{2}\ln y + \ln(1+\sqrt{1+y^{-3}}), \quad (12)$$

where $y = \frac{a}{a_\star}$ and $t_i$ represent the initial time when $a(t_i) = 0$. Now we have two conditions for cosmic time evolution of the universe. At early time of the universe $y \ll 1$ which leads the RHS of the Eq. (12), i.e. $\frac{4\pi G\alpha(t-t_i)}{\gamma} \approx 2y^{\frac{3}{2}}$, and for late time $y \gg 1$ and we have left with $\frac{4\pi G\alpha(t-t_i)}{\gamma} \approx \frac{3}{2}\ln y$. Variation of the cosmic time evolution of Eq. (12) with respect cosmic scale factor ($a$) has been shown in Fig. 2. The figure clearly indicating that the cosmic time evolution of the universe from the early decelerating phase to present accelerated phase.

### 2.2 Energy density

Now using Eq. (4) along with Eq. (2) one can obtain the energy density of the dark energy as,

$$\rho_D = \frac{4\pi G\alpha^2}{3\gamma}\left(1+\sqrt{1+\left(\frac{a_\star}{a}\right)^3}\right)$$
$$\left[1+\frac{4\pi G\beta}{3\gamma}\left(1+\sqrt{1+\left(\frac{a_\star}{a}\right)^3}\right)\right]. \quad (13)$$

Variation of $\rho_D$ with respect to scale factor $a$ is shown in Fig. 3. From the plot it is clear that the density of the universe is decreasing with the expansion of the universe.

### 2.3 EoS parameter

The equation of state ($\omega_D$) parameter for generalized ghost dark energy can be obtained from the following equation as:

$$\omega_D = \left(\frac{a_\star}{a}\right)^3 \left(\frac{1}{\sqrt{1+\left(\frac{a_\star}{a}\right)^3}} - \frac{1}{1+\sqrt{1+\left(\frac{a_\star}{a}\right)^3}}\right)$$
$$\left(\frac{1+(1-\gamma)\sqrt{1+\left(\frac{a_\star}{a}\right)^3}}{2+(1-\gamma)\left(\sqrt{1+\left(\frac{a_\star}{a}\right)^3}-1\right)}\right) - 1$$
$$= \begin{cases} 0, & a \ll a_\star \\ -1, & a \gg a_\star. \end{cases} \quad (14)$$

From this values of $\omega_D$ we can conclude that at late time the dark energy behaved like a cosmology constant as $\omega_D$ is asymptotic. In Fig. 4 we plot the graph of $\omega_D \sim a$. From the figure, we see that dark energy is quintessence as $\omega_D$ can never cross -1. $\omega_D$ varies from zero at early time to -1 at late time. Now we rewrite the expression of $\omega_D$ as

$$3(1+\omega_D) = -\frac{\dot{H}}{H^2}\left(\frac{\alpha+2\beta H}{\alpha+\beta H}\right) = H^{-1}f(H). \quad (15)$$

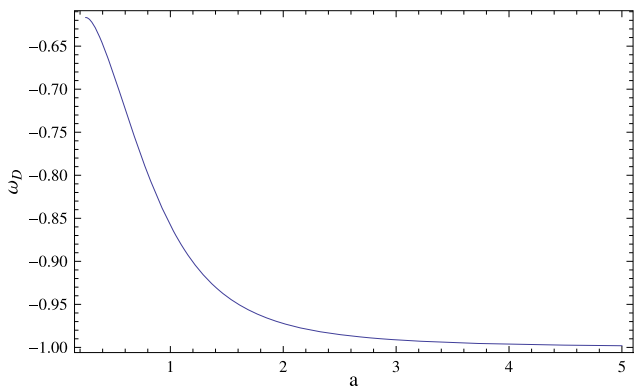

**Fig. 4** Variation of EOS parameter with $a$

From this expression, we see that the Equation of state of GGDE tightly relates to variation of Hubble parameter. Also $H^{-1}\dot{f}(H) \sim 3$ in dust phase and $H^{-1}\dot{f}(H) \sim 0$ in the de Sitter phase which imply that Hubble is different in different phases of Universe. So in transit of Universe from the dust phase to the de Sitter phase there will be a jump from 0 to $-1$.

To understand the fitting result in section III, we calculate the equation of state in GGDE model. Then the value of $\omega_{D0}$ at present time is-

$$\omega_{D0}(a=1) = -\frac{\alpha}{2(\alpha+\beta H_0) - \Omega_{D0}(\alpha+2\beta H_0)}$$
$$= -\frac{\alpha}{\alpha + \Omega_{DM0}(\alpha+2\beta H_0)}. \quad (16)$$

From Raychaudhuri Eq. (6), the total equation of state of the universe is given below-

$$\omega_{Dtot} = -1 + \frac{2(\alpha+\beta H)}{\alpha+2\beta H}(1+\omega_D)$$
$$= \begin{cases} 0, & a \ll a_\star \\ -1, & a \gg a_\star. \end{cases} \quad (17)$$

This implies that the expansion of Universe switches from deceleration to acceleration with the value of $\omega_{Dtot}$ which is monotonically decreases from 0 at early time to $-1$ at late time.

### 2.4 Model

In this section, we will include baryon and radiation in flat FRW universe to fit the model with observational data. Then the Friedman equation reads as,

$$H^2 = \frac{8\pi G}{3}(\alpha H + \beta H^2 + \rho_m + \rho_b + \rho_r). \quad (18)$$

We rewrite the equation as-

$$\Psi \equiv \frac{H}{H_0} = \frac{B}{2\gamma} \pm \frac{1}{\gamma}$$
$$\sqrt{\frac{B^2}{4} + \gamma\left((\Omega_{m0}+\Omega_{b0})(1+z)^3 + \Omega_{r0}(1+z)^4\right)}, \quad (19)$$

where $B = \Omega_{D0} - \frac{8\pi G\beta}{3}$, $\Omega_{r0}$ and $\Omega_{b0}$ are the present values of dimensionless energy density for baryon and radiation respectively. The energy density of CDM and baryon are written as $\Omega_{m0} + \Omega_{b0} = \Omega_{DM0}$. As we explain the acceleration in flat Universe. So we notice that we get $\Omega_{D0} + \Omega_{r0} + \Omega_{DM0} = 1$. We know that, the sum of photons an relativistic neutrinos is the energy density of radiation $\Omega_{r0} = \Omega_{\nu 0}(1 + 0.2271N)$ where the effective number of neutrino species $N = 3.04$ and $\Omega_{\nu 0} = 2.469 \times 10^{-5} h^{-2}$ for $T_{CMB} = 2.725\ K$ ($h = \frac{H_0}{100}\ Mpc\ km\ s^{-1}$). Using the definition of dimensionless energy density of dark energy and flatness of our Universe we can notice an important relation as-

$$\left[(1-\Omega_{DM0}) - \frac{8\pi G\beta}{3}\right]H_0 = \frac{8\pi G\alpha}{3} = constant. \quad (20)$$

The value of $\Omega_{\nu 0}$ is negligible compare to the value of $\Omega_{DM0}$. From Eq. (20) we observed that the parameters $\Omega_{DM0}, \beta, h$ and $\alpha$ are closely related. So there exist an relation between $\alpha$ and $\beta$ in terms of $\Omega_{DM0}$ and $h$. Now we choose $\Omega_{DM0}, h$ and $\Omega_{b0}$ as free parameter of the model in this analysis.

### 2.5 Analysis with stern data sets

Observational data analysis is one of the most important tool to study the viability of various physical parameters of any model. In this section we are going to investigate the constraints on the model parameters obtained from the observational data fitting. These model parameters can be determined by $H(z) - z$ (Stern) data analysis [50–55]. The mechanism that we have adopted for analysing the observational data is the minimization of $\chi^2$, here we take the minimization of $\chi^2$ for the Hubble-redshift Stern data set [56] that yields

$$\chi^2_{H-z} = \sum \frac{(H(z) - H_{obs}(z))^2}{\sigma_z^2}, \quad (21)$$

where $H(z)$ and $H_{obs}$ respectively represent the theoretical and observational values of the Hubble parameter for our model at different redshifts ($z$) and $\sigma_z$ represents the error in relation with the observation.

### 2.6 Fitting results

First, we have obtained the fitting values of the model parameters $\alpha$ and $\beta$ using the values of $\Omega_{m0}$ and $H_0$ from Plank 2015 results [49]. Then we have used these values of $\alpha$ and $\beta$ to obtain the data fitting result of $\Omega_{DM0}$ and $h$ from Stern data Sets.

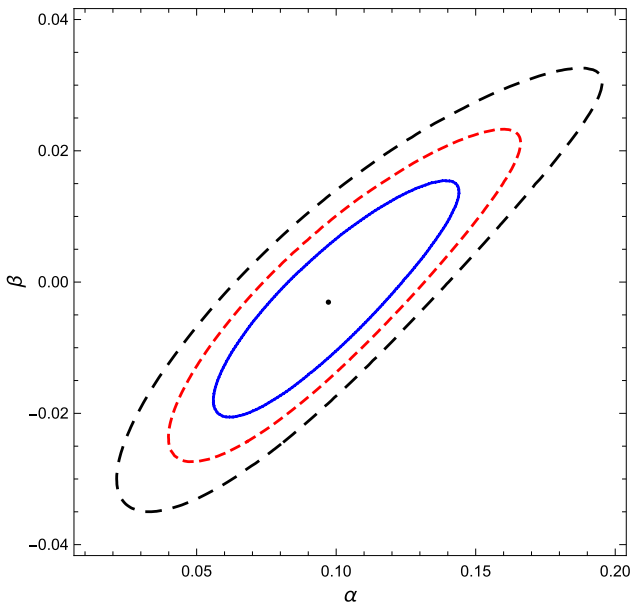

**Fig. 5** Contour plot of $\alpha$ and $\beta$

**Table 1** The best-fit values for the model parameters $\alpha$, $\beta$ and $\gamma$

| Parameter | $\alpha$ | $\beta$ | $\gamma$ |
|---|---|---|---|
| Best-fit | 0.01 | $-0.003$ | 1.025 |

**Table 2** The best-fit values with $1\sigma$ and $2\sigma$ errors for $\Omega_{mo}$, $h$ and $\Omega_{bo}$ in the generalized ghost dark energy model

| Parameter | $\Omega_{DM0}$ | $h$ |
|---|---|---|
| Best-fit $^{-1\sigma,-2\sigma}_{+1\sigma,+2\sigma}$ | $0.210^{-0.077,-0.102}_{+0.089,+0.121}$ | $0.689^{-0.065,-0.095}_{+0.064,+0.089}$ |

**Table 3** The best-fit values for the $\Lambda CDM$ model using the same data set

| Parameter | $\Omega_{DM0}$ | $h$ |
|---|---|---|
| Best-fit | 0.273 | 0.703 |

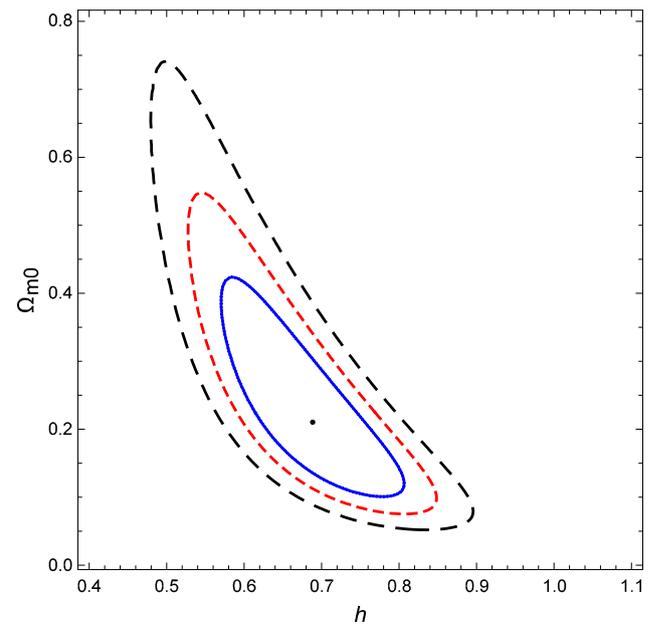

**Fig. 6** Contour plot in $\Omega_{DM0}$-$h$ Plane

We have shown the contour plot between the parameters $\alpha$ and $\beta$ in Fig. 5. Again from Table 1, we have obtained the values of the parameters as $\alpha \approx 0.01$ and $\beta \approx -0.003$. Using these values of model parameters we have also calculated $\gamma \approx 1.025$ for planck observation results of 2015, whereas in [40] the value of $\gamma$ has been obtained as $\gamma \approx 1.1$, i.e., our results can be treated as a close agreement with the results in the Ref. [40].

In Table 2, we have summarized the best fit values and errors of parameter. For comparison these values with $\Lambda CDM$ model we list the best fit values of corresponding parameters in Table 3. From these two tables we can see that our best fit result are slightly smaller than corresponding values in $\Lambda CDM$.

Using the best fit values of $\alpha$ and $\beta$ from contour plot in Fig. 5 we have obtained the results of the different physical parameters of the present article and it has been observed that the results are physically acceptable.

Now from the two dimensional contour plot shown in Fig. 6, we can observe that there exist a strong co-relation between $\Omega_{DM0}$ and $h$.

There is a shifting from the dust phase to the de Sitter phase at $a_\star = 1.236$ with $\Omega_{DM0} = 0.210$. So the Universe's acceleration begin $t_{acc} = \frac{a_\star}{2} = 0.618$ or in term of redshift $z_{ace} = 0.617$. and the present values of EoS is $\omega_{D0} = -0.83$.

## 3 Classical stability analysis

In order to check the classical stability of our present model we have studied the behaviour of deceleration parameter ($q$) and adiabatic sound speed ($v_s^2$). For any stable solution $v_s^2$ must have positive. To analysed the stability we have used the results obtained from the best fitting in the previous section.

### 3.1 Deceleration parameter

To measure the cosmic acceleration of the expansion of space in FRW universe one can study deceleration parameter, a dimensionless parameter which can be defined as-

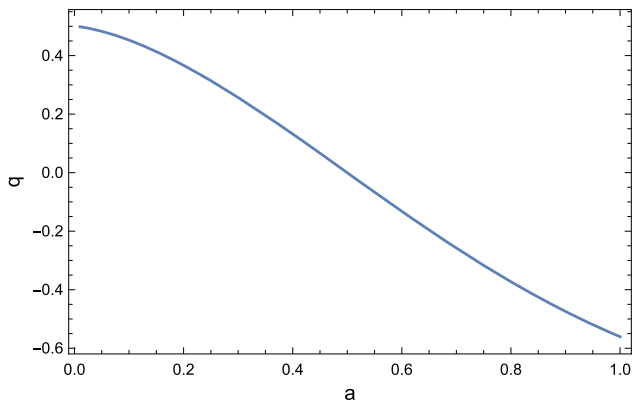

Fig. 7 Variation of Deceleration parameter with $a$

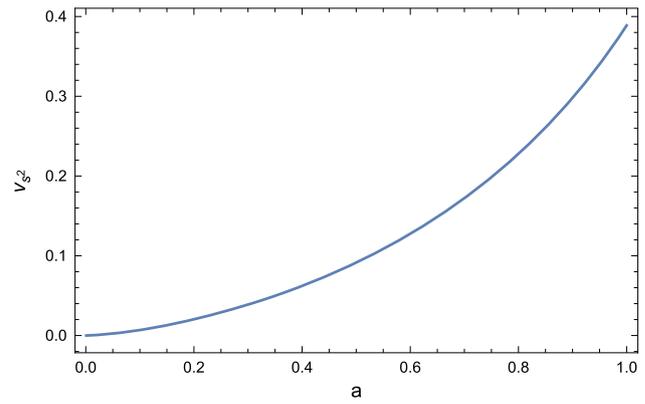

Fig. 8 Variation of speed of sound with $a$

$$q = -\frac{\ddot{a}a}{\dot{a}^2} = -\left(1 + \frac{\dot{H}}{H^2}\right), \qquad (22)$$

where $\ddot{a}$ represent the second order time derivative of the scale factor. Now putting corresponding values of $H$ and $\dot{H}$ from Eq. (8) in the above equation eventually we have

$$q = -1 + \frac{3}{2}\left(\frac{a_\star}{a}\right)^3 \left[\frac{1}{\sqrt{1+\left(\frac{a_\star}{a}\right)^3}} - \frac{1}{1+\sqrt{1+\left(\frac{a_\star}{a}\right)^3}}\right]$$
$$= \begin{cases} \frac{1}{2}, & a \ll a_\star \\ -1, & a \gg a_\star. \end{cases} \qquad (23)$$

At the early epoch where $\Omega_D \to 0$ the value of deceleration parameter has the value $q = \frac{1}{2}$, whereas in late time where the universe is dominated by dark energy it becomes $q = -1$. This indicates that our universe undergo a transition from deceleration phase at early time to acceleration phase at present time. We have shown the variation of the deceleration parameter in Fig. 7 that shows the transition of deceleration phase to the acceleration phase at $a \approx 0.52$ of the universe.

### 3.2 Adiabatic sound speed

We now study the adiabatic sound speed to get the classical stability of the GGDE model. The squared sound speed has been found for model is as follows [32]

$$v_s^2 = \frac{\dot{p}_D}{\dot{\rho}_D} = \frac{2\alpha\rho_{DM_0}\gamma}{\frac{8\pi G\alpha^2 a^2}{3} + 4\rho_{DM_0}\gamma A}, \qquad (24)$$

where $A = \alpha + \beta\sqrt{\frac{\frac{8\pi G}{3}\left(\frac{8\pi G\alpha^2 a^3}{3} + 4\rho_{DM_0}\gamma\right)}{a^3}}$.

Variation of the squared sound speed with respect to the scale parameter has been shown in Fig. 8. From the figure we can observe that squared speed of sound is always positive and less than unity throughout the evolution of the universe. Which suggests the classical stability of our model. Cai et al. [40] have claimed that for the best fitting results the coefficient of the sub-leading term must be negative. We have found that for negative value of $\beta$, the model shows classical stability. The above result is true in the case for GGDE model as there is no any propagating degrees of freedom (see Ref. [42]). We get the value of sound speed within 1 ($v_s^2 < 1$) throughout the stellar evolution.

Now one can reproduce the results obtained by Cai et al. [32] by setting $\beta = 0$ in Eq. (24), i.e., GDE model. We have showed the variation of squared sound speed ($v_s^2$) with respect to scale factor $a$ which remains negative throughout the evolution of the universe for GDE model, this clearly indicates the instability of the universe against perturbations under the GDE background. Moreover in the GGDE model the variation of squared sound speed ($v_s^2$) with respect to scale factor $a$ remains positive, i.e., contribution of the sub-leading term in Eq. (2) of GGDE provides a stable solution of the universe. So $\beta$ has the substantial contributions to get classically stability of the model.

In order to obtain an insight on the stability issue of the GDE in FRW universe by setting $\beta = 0$ in the relation Eq. (24). Using this value we plot the figure of sound speed with respect to the scale parameter for ghost dark energy has been shown in Fig. 9. Which clearly indicates that the squared sound speed remains negative indicating the instability of the universe against perturbations in the GDE background. This recover the result studied by Cai et al. [32].

## 4 Physical results

In this article we have studied the generalized ghost dark energy (GGDE) under the framework of flat homogenous and isotropic FRW universe. Contribution of the subleading term of Eq. (2) leads to a classically stable and physically acceptable solution of the universe. Using the values obtained from the best fitting results of the model parameters as shown in Fig. 5 and Table 1, we have studied the variation of various physical parameters of GGDE in this article. Again the value of $\beta$ has been found to be negative which indicates that subleading term has negative contribution to the energy density. In this section we are going to discuss some of the important features that we have analysed here.

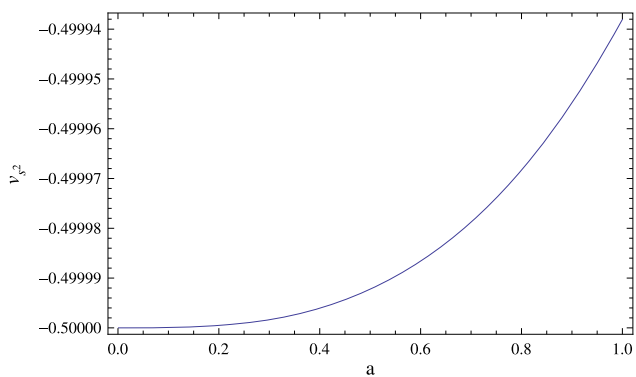

**Fig. 9** Variation of speed of sound for GDE with $a$

(i) We have solved the noninteracting Friedman equation of Eq. (4) to obtain Hubble parameter for our model. Variation of the Hubble parameter has been shown in Fig. 1, value of the Hubble parameter is decreasing with the scale factor. From the results of data fitting we get the value of $H \approx 68.9$, which is an well agreement with the present observational value.

(ii) Using Eq. (13) we have plotted the variation of energy density of GGDE for the universe with respect to cosmic scale factor $a$ in Fig. 3. It clearly indicates that with the increase of scale factor the energy density decreases. Present observation results have also claimed that the energy density become lesser due to the expansion of the universe.

(iii) EoS parameter: From the observational results it have been confirmed that our universe has experienced a phase of expansion with an acceleration due to the influence of dark energy. Thus these new models are capable to provide stable solutions of the universe dominated by DE. From the variation of Equation of state parameter in Fig. 4, it has been observed that $\omega_D$ for GGDE model always lies within the Phantom line $\omega_D = -1$ and at late time it asymptotically approaches to $-1$.

(iv) Deceleration parameter: We have calculated the values of deceleration parameter for our model in Eq. (23) and its variation has been shown in Fig. 7. At early epoch of the universe when $\Omega_D \ll 1$ the value of deceleration parameter $q = \frac{1}{2}$, which indicates that at that stage the universe was evolved due to domination of the DM component, i.e., the universe was experiencing a phase of deceleration. For late time this phase of deceleration changed to acceleration due to presence of dark energy. From the variation in Fig. 7 we observe that flipping of signature of the deceleration parameter from early epoch to late time, which exactly matches with the observational evidences.

(v) Sound speed: The expression for adiabatic sound speed ($v_s^2$) has been obtained in Eq. (24), it shows that $v_s^2$ is positive. Using the best fitting values of the model parameters $\alpha$ and $\beta$ we have plotted the variation of sound speed with respect to scale parameter ($a$) in Fig. 8 and it can be found that sound speed remains positive and always less than 1. We have also showed the variation of $v_s^2$ for $\beta = 0$, which shows that $v_s^2$ is negative through out the evolution. This comparative study of sound speed confirms the classical stability of GGDE model over the GDE model [40].

## 5 Conclusions

The present work can be claimed as a continuation of previous works related to generalized ghost dark energy [40]. Most of the earlier works were unable to provide a stable solution of GDE due to negative value of adiabatic sound speed. Cai et al. [32] gave a clear indication that for stable solution the subleadng term must have negative contribution in the energy density for GGDE model of Eq. (2). After fine tuning of beta we are able to obtain a classical stable solution of the universe considering GGDE model. With this value we have obtained the solutions for various features, all of them clearly indicating the classical stability as well as the acceptability of our present model on Generalized Ghost Dark energy.

**Acknowledgements** UD is thankful to the Inter-University Centre for Astronomy and Astrophysics (IUCAA), Pune, India to provide Visiting Associateship under which a part of this work was carried out. MB and SG also thankful to the Inter-University Centre for Astronomy and Astrophysics (IUCAA), Pune, India for providing the working facilities and hospitality during visit.

**Data Availability Statement** This manuscript has no associated data or the data will not be deposited. [Authors' comment: This manuscript has no associated data so the data will not be deposited.]